# Laboratory studies on the viability of life in $H_2$-dominated exoplanet atmospheres

S. Seager[1,2,3], J. Huang[4], J.J. Petkowski[1], M. Pajusalu[1]
[1]Department of Earth, Planetary, and Atmospheric Sciences, Massachusetts Institute of Technology
[2]Department of Physics, Massachusetts Institute of Technology
[3]Department of Aeronautics and Astronautics, Massachusetts Institute of Technology
[4]Department of Chemistry, Massachusetts Institute of Technology

**Theory and observation for the search for life on exoplanets via atmospheric "biosignature gases" is accelerating, motivated by the capabilities of the next generation of space- and ground-based telescopes. The most observationally accessible rocky planet atmospheres are those dominated by molecular hydrogen gas, because the low density of $H_2$-gas leads to an expansive atmosphere. The capability of life to withstand such exotic environments, however, has not been tested in this context. We demonstrate that single-celled microorganisms (*E. coli* and yeast) that normally do not inhabit $H_2$-dominated environments can survive and grow in a 100% $H_2$ atmosphere. We also describe the astonishing diversity of dozens of different gases produced by *E. coli*, including many already proposed as potential biosignature gases (e.g., nitrous oxide, ammonia, methanethiol, dimethylsulfide, carbonyl sulfide, and isoprene). This work demonstrates the utility of lab experiments to better identify which kinds of alien environments can host some form of possibly detectable life.**

Introduction
There are several ideas of how rocky exoplanets (small planets with radii smaller than about 1.7 Earth radii) may have formed and maintained $H_2$-dominated atmospheres. Rocky exoplanets release $H_2$ gas to their atmospheres as water reacts with metallic Fe in accreting materials during planet formation[1,2]. A planet that accreted from iron-rich primitive material (e.g., similar to EH chondritic meteorites) and water ice may have an $H_2$-dominated atmosphere up to a few percent of the total planet mass, if all the iron and water reacted[2,3]. This extreme end member scenario provides rationale that planets may have an $H_2$-dominated atmosphere even after losing much of the envelope from stellar-EUV radiation, or even if smaller than maximally possible amounts $H_2$ outgassed during planet accretion. Furthermore, super Earths more massive (with higher surface gravity) and colder than Earth may maintain an $H_2$ atmosphere against thermal Jeans escape[4]. A mechanism for replenishing an $H_2$ atmosphere against atmospheric escape may occur in super Earths with predominantly icy interiors: the high pressures convert methane ice into ethane, butane, or even elemental carbon, continually releasing $H_2$[5].

Super Earth exoplanets may have captured a $H_2$/He atmosphere from the protoplanetary disk, in contrast to planets that formed an $H_2$ atmosphere from outgassing. Planets of a few Earth masses beyond about 2 AU from their host star, away from destructive host star XUV radiation may maintain a primordial $H_2$-He atmosphere of 1 to100 bars, provided the planet has protective magnetic fields[6]. Note that at orbital separations of 2 AU or greater, the planet will still be habitable as $H_2$ acts as a powerful greenhouse gas[6], due to collision-induced infrared $H_2$ opacity. Slow preferential escape of H in special cases may lead to a secondary He atmospheres after billions of years[7]. Although not yet observable, rogue rocky planets ejected from their planetary systems can also retain massive primordial $H_2$ atmospheres[8]. Despite very cold outer layers, the surface could be clement from interior radioactive heat trapped by the $H_2$ atmosphere greenhouse. Note that in this work we focus on atmospheres, which are distinct from more massive gas envelopes found on Neptune- and sub-Neptune size exoplanets.

It is known that Earth had very small amounts of $H_2$ in its early atmosphere, up to about 0.1%[9] billions of years ago. The $H_2$ persisted for hundreds of millions of years[9] or possibly even for 2.5 billion years up to the Great Oxidation Event[10]. Today, the little $H_2$ produced on Earth is consumed by microorganisms, oxidized in the atmosphere, or lost to space.

Initial attempts to observe rocky exoplanet atmospheres of any kind are limited by current telescope capabilities. The few observed with the most capable instrument, the *Hubble Space Telescope* WFC3 have not shown evidence for $H_2$-dominated atmospheres (e.g., Trappist-1 d, e, f, g[11] and LHS 1132b[12]), although haze- or cloud-covered $H_2$-dominated atmospheres are not ruled out by the observations. Yet, exoplanet atmosphere studies via transmission spectroscopy[13] and direct imaging[14] are solidly established; dozens of hot giant exoplanet atmospheres have been successfully observed with detection of gases including Na, K, CO, $H_2O$, $CH_4$, TiO and inferences of clouds or hazes[15,16]. Astronomers are therefore confident small exoplanet atmospheres for planets orbiting small red dwarf stars can be observed via transmission spectra with the NASA-ESA *James Webb Space Telescope* (JWST) and via direct imaging the extremely large ground-based telescopes now under construction. The goal in observing rocky planet atmospheres includes the identification of greenhouse gases to estimate exoplanet surface temperature, the search for water vapor, indicative of surface liquid water needed for all life as we know it, and the search for "biosignature" gases that might attributed to life.

Rocky exoplanets with $H_2$ atmospheres will be far easier to detect and study than those with atmospheres composed of higher mean molecular weight gases such as $CO_2$ and $N_2$. Atmospheric pressure (and density) falls off exponentially with increasing altitude from the planetary surface, with an *e*-folding factor of "scale height", $H = kT/\mu m_H g$. Here $k$ is Boltzmann's constant, $T$ is temperature, $g$ is surface gravity and $\mu$ is the mean molecular weight, and $m_H$ is the atomic mass unit (roughly the mass of a hydrogen atom). For atmospheres observed in transmission or reflection (but not thermal emission), a scale of about a few times $H$ is the relevant factor for exoplanet observation spectral signatures. Based on the scale height estimate, an $H_2$ atmosphere is 14 times larger in extent than a $N_2$-dominated atmosphere.

Given the facts that: rocky exoplanets with $H_2$-dominated atmospheres likely exist; such planet atmospheres are more easily observed than $N_2$- or $CO_2$-dominated atmospheres; and the next generation telescopes with the capability to study rocky planet atmospheres are coming online in the next several years, it is important to assess planets with $H_2$-dominated atmospheres for viability of life.

We are building on limited past work. Only a small set of simple microorganisms that are normally dependent on H$_2$ to survive (and are thus accommodated to H$_2$) have been studied in high H$_2$ gas concentration environments. Methanogens (e.g.,[17]) (and acetogens[18]) are routinely grown in 80% H$_2$ and 20% CO$_2$. Very few other groups of microorganisms have been studied in high H$_2$ gas concentrations (e.g., sulfate reducing bacteria, and hydrogen oxidizing bacteria, such as knallgas bacteria). Eukaryotes have not been studied in high concentration H$_2$-environments (except in a couple of isolated cases in passing) and no studies exist of yeast in high H$_2$-environments. Although H$_2$ is not known to be toxic to life in either small or large quantities (negative health effects for animals are associated only with asphyxiation through displacement of O$_2$ in the lungs), microorganisms have not been shown to grow in pure 100% H$_2$ atmospheres before.

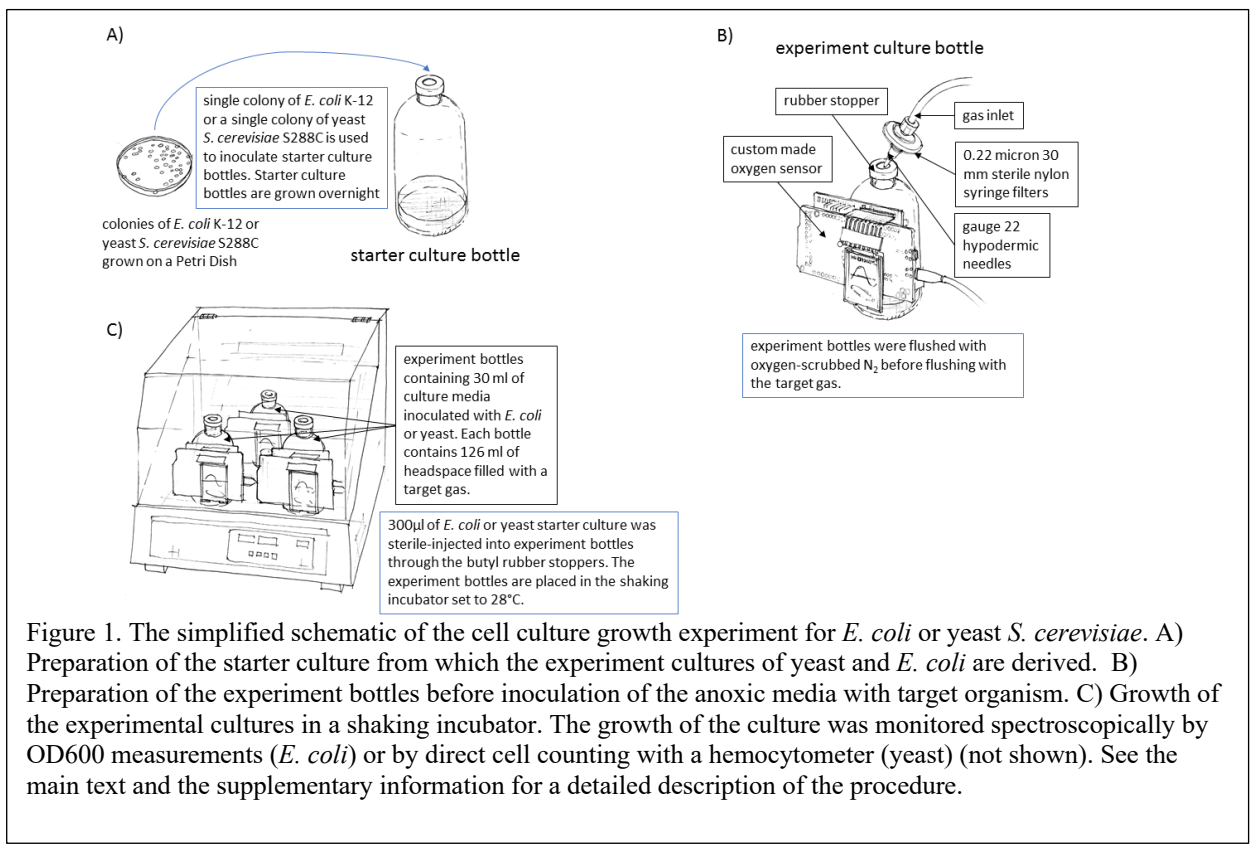

Figure 1. The simplified schematic of the cell culture growth experiment for *E. coli* or yeast *S. cerevisiae*. A) Preparation of the starter culture from which the experiment cultures of yeast and *E. coli* are derived. B) Preparation of the experiment bottles before inoculation of the anoxic media with target organism. C) Growth of the experimental cultures in a shaking incubator. The growth of the culture was monitored spectroscopically by OD600 measurements (*E. coli*) or by direct cell counting with a hemocytometer (yeast) (not shown). See the main text and the supplementary information for a detailed description of the procedure.

**Approach**

We conducted growth experiments (Figure 1; Supplementary Figure 1) on two species of microorganisms *Escherichia coli* strain K-12 and yeast *Saccharomyces cerevisiae* strain S288C in our custom-built bioreactor system. The system consists of small borosilicate bottles with 30 mL of culture media (standard media for *E. coli* and yeast cell culture growth) and 126 mL of headspace. The headspace each bottle was flushed with appropriate gases via a needle injected through a sterile rubber stopper. The anaerobic experiments (100% H$_2$, 100% He, and 80% N$_2$ and 20% CO$_2$) were started in serum bottles that had been previously flushed with oxygen-scrubbed N$_2$ before flushing with the target gas. Four bottles (100% H$_2$, air, 100% He, 20% CO$_2$ and 80% $N_2$) were placed in an incubator shaker at 28 °C. We ensured the gas concentration remained stable and the cultures remained anoxic over the duration of the experiment with micromolar $O_2$ levels via continuous measurements using our new, precise and accurate $O_2$ sensors[19] (Supplementary Figure 2; Supplementary Figure 3). The bottles were continuously monitored for a quick assessment of cell culture turbidity by a custom camera. We sampled the culture periodically to assess growth of the culture (*E. coli* by optical density measurements (OD600) and yeast by cell counting with a hemocytometer). See Supplementary Information for more information on approach and methods.

Because of equilibrium gas exchange, the liquid medium will be saturated with $H_2$. The solubility of $H_2$ in water is lower than $N_2$ and $O_2$ (but of the same order of magnitude), with $O_2$ being the most soluble of the three gases (approximately 1.6 times more soluble than $H_2$)[20]. We monitor $O_2$ partial pressure with custom sensors[19] to show that the $O_2$ stays at trace levels for the duration of the experiments. The gas-phase partial pressures of $O_2$ are less than 60 μbar for the *E. coli* experiment and down to a few microbars for the yeast experiment (Supplementary Information), translating into about 70 and 2 nM of dissolved oxygen concentration in water, respectively. For context, these $O_2$ gas concentration values are close to the upper limits of found on Archean Earth[21], before Earth's atmosphere was oxygenated.

## Results

We consider a pure 100% $H_2$ atmosphere as a control; if life can survive in a 100% $H_2$ atmosphere then it can also survive in an $H_2$-dominated atmosphere. We show that a 100% $H_2$ atmosphere has no detrimental effects on microorganisms that do not normally inhabit $H_2$-rich conditions. We chose *E. coli* and yeast because they are the standard model organisms used in biology. *E. coli* as a representative of the domain Bacteria and yeast for Eukarya. We show that both *E. coli*, a simple single-celled prokaryote as well as yeast, a more complex single celled eukaryote can survive and reproduce in a liquid cultures surrounded by a 100% $H_2$ gas environment.

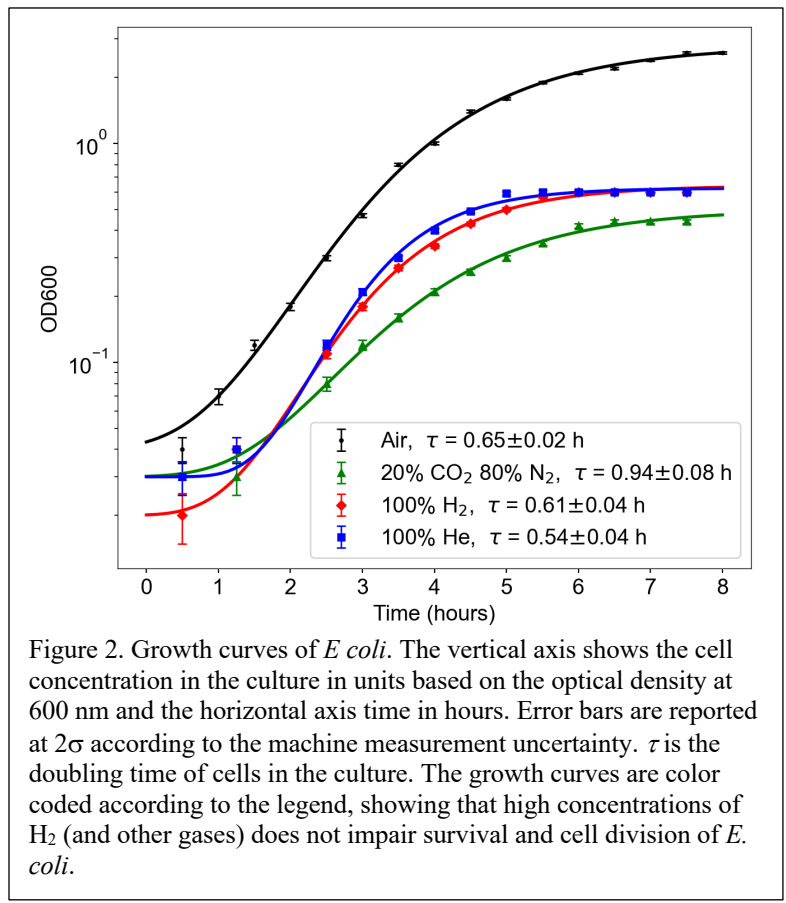

Figure 2. Growth curves of *E coli*. The vertical axis shows the cell concentration in the culture in units based on the optical density at 600 nm and the horizontal axis time in hours. Error bars are reported at 2σ according to the machine measurement uncertainty. $\tau$ is the doubling time of cells in the culture. The growth curves are color coded according to the legend, showing that high concentrations of $H_2$ (and other gases) does not impair survival and cell division of *E. coli*.

The growth curves (Figures 2 and 3, for *E. coli* and yeast, respectively) demonstrate that the organisms are reproducing normally. For *E. coli* grown in a 100% $H_2$ atmosphere, the maximal cell concentration (number of cells per unit volume) is just slightly over two times less than in the control of *E. coli* grown in air. If the availability of $O_2$ is low, *E. coli* switches from aerobic respiration to the less efficient energy metabolism (smaller amount of energy produced per unit of catabolized organic material) based on either anaerobic respiration or fermentation, but the experimental data shows near identical maximal growth rates. We note that *E. coli* (and yeast) derives energy from materials in the liquid culture medium. Other controls of a pure He gas environment show similar *E. coli* growth curves as compared to the $H_2$ gas environment, for the same reasons. The control containing a 20% $CO_2$/80% $N_2$ gas mixture shows demonstrably slower growth compared to the $H_2$ and He gas environments, likely because dissolved $CO_2$ makes the medium more acidic, slowing growth, although this does not seem to have an effect on maximum cell concentration.

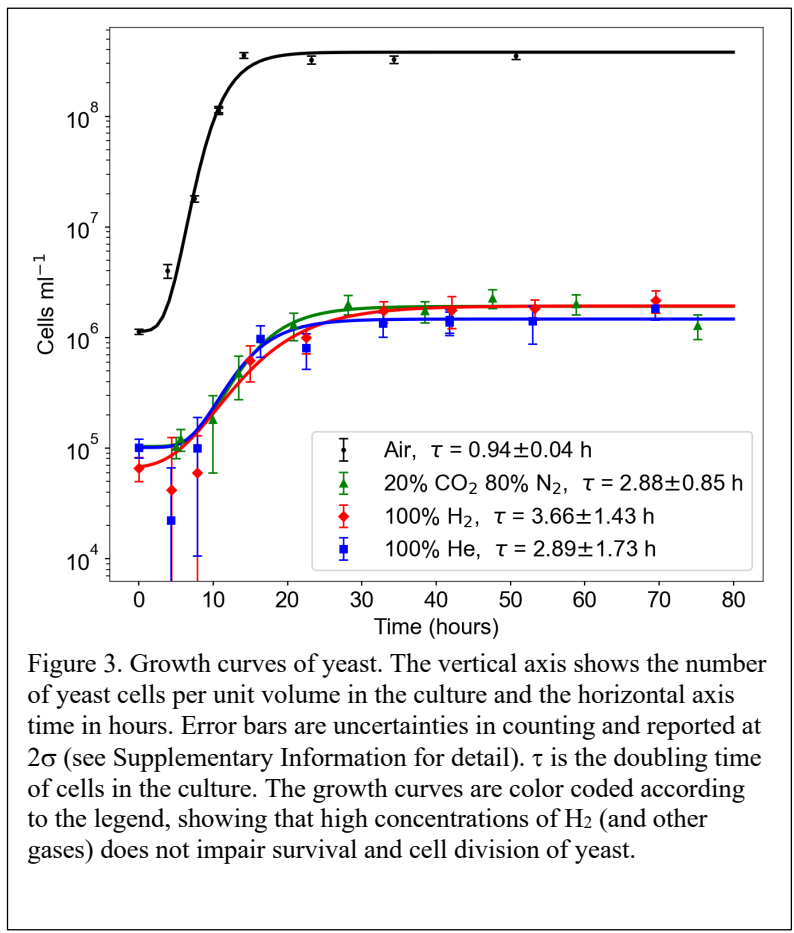

Figure 3. Growth curves of yeast. The vertical axis shows the number of yeast cells per unit volume in the culture and the horizontal axis time in hours. Error bars are uncertainties in counting and reported at $2\sigma$ (see Supplementary Information for detail). $\tau$ is the doubling time of cells in the culture. The growth curves are color coded according to the legend, showing that high concentrations of $H_2$ (and other gases) does not impair survival and cell division of yeast.

Yeast has a substantially lower maximal cell concentration in the pure $H_2$ environment (and in other anaerobic cases), two and a half orders of magnitude lower than for the control experiment of yeast growing in air, accompanied by a roughly three times longer generation time. Yeast growth results in the controls of He and 20% $CO_2$/80% $N_2$ are within uncertainties identical to those of the pure $H_2$ environment, with no distinction for the 20% $CO_2$/80% $N_2$. Yeast is not affected by the acidification of the medium as yeast is known to grow in moderately acidic media with pH as low as 4[22].

We explain the disproportionately low growth rates and maximal cell densities of yeast as due to the absence of biochemically relevant $O_2$, which is unrelated to energy metabolism. Independent of aerobic respiration[23], $O_2$ is a crucial substrate in the biosynthesis of many biochemicals essential to different eukaryotes. For example, unsaturated fatty acids, quinones, porphyrin rings of hemes and chlorophylls, and sterols. For yeast specifically, it is well known that $O_2$ is required as a biosynthetic substrate for several critical metabolites (e.g. heme and sterols, like ergosterol). Under anaerobic conditions, yeast cells start to rely exclusively on an import of exogenous sterols from the environment[24]. Thus, the lack of availability of sterols in the culture medium is likely the main limiting growth factor in the very low $O_2$ environment of our experimental conditions. The exceptionally slow growth rates for yeast should not apply to all eukaryotes. While yeasts are obligatorily dependent on sterols for growth, other eukaryotes (that exclusively inhabit $O_2$-free environments) substitute sterols with other (sterol-like) molecules that do not require $O_2$ as a substrate for synthesis[25].

**Discussion**

Pockets of high $H_2$ concentration (from, e.g., Ca decay, serpentinization) exist on modern Earth, including microniches in mines with up to 30 to 88% $H_2$ by volume[26]. These $H_2$-rich environments are also populated by microbes (such as sulfur reducing bacteria and some archaea). Life on Earth uses $H_2$ for chemoautotrophy, for anaerobic respiration. It is unknown if more complex eukaryotic microorganisms are inhabiting such high $H_2$ niches, but our experiments support the idea that in principle they could be.

An atmosphere with $H_2$ is likely to be accompanied by methane, $CH_4$. Methane will form in the atmosphere at moderate temperatures if four times as much hydrogen as carbon is present[27]. Also, along with $H_2$, $CH_4$ may outgas gradually from an initial large interior $H_2O$ reservoir to replenish volatiles in the atmosphere[28]. $CH_4$ gas is non-toxic in the sense that $CH_4$ does not readily chemically react in the detrimental fashion with the biochemicals of the cell. $CH_4$'s negative biological effects, as for $H_2$ and $N_2$, would be associated only with displacement of $O_2$ and creating too low $O_2$ partial pressures for aerobic organisms[29].

An $H_2$-dominated atmosphere with its trace amounts of reduced gases is conducive to the origin of life, not detrimental, because reduced precursor molecules are presumed to be needed for life's origin (for example, see ref [30]). For example, important organic precursor molecules (such as nitriles or carbonyls including aldehydes, ketones and amides, etc.) that could eventually participate in the formation of biologically important molecules like nucleotides (e.g. in RNA and DNA) or amino acids (e.g. in proteins) can form much more easily if reduced gases are present than if they are not. This is especially relevant if life originated on the surface of the planet (for example, see ref[31]) and not at deep sea vents.

As an example of biosignature gases that could be present in an $H_2$-dominated atmosphere, we can consider gases produced by *E. coli*. *E. coli* synthesizes an impressive number of volatile molecules (45) with a wide variety of functional groups (10; Supplementary Table 1; Supplementary Table 2). Several *E. coli*-produced gases have already been studied as promising exoplanet biosignature gases (including ammonia, methanethiol, dimethylsulfide, carbonyl sulfide, carbonyl disulfide, nitrous oxide, isoprene, and possibly-produced by *E. coli* methane and phosphine; Supplementary Table 1) and have relatively distinctive spectral features from each other (Figure 4). Yeast produces even more gases than *E. coli* (75 vs. 45), but the gases fall into fewer functional group categories (Supplementary Table 2). Many of the *E. coli*- or yeast-produced gases have yet to be evaluated as biosignatures, including a large number of challenging-to-spectrally-distinguish carbonyls and alcohols.

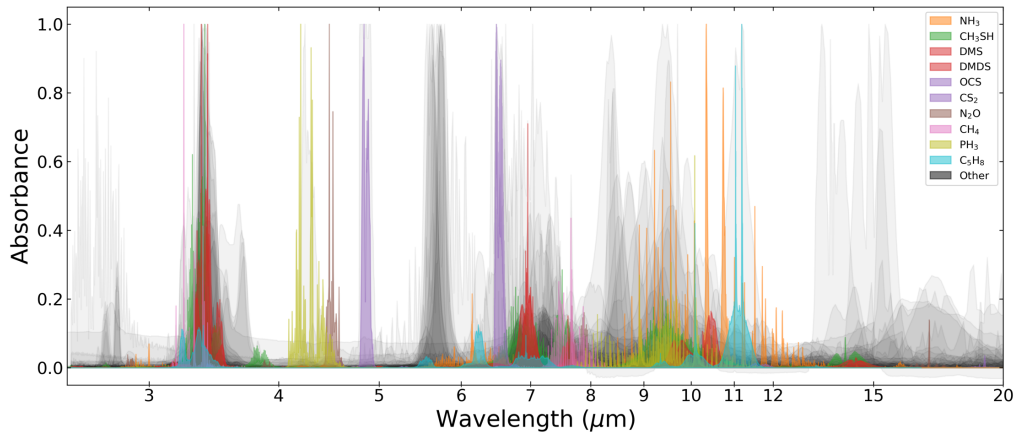

Figure 4. Spectral features of gases produced by *E. coli* with existing data. Absorbance (normalized to 1) vs. wavelength in microns. Molecules are as noted in the legend. Note that $CH_4$ and $PH_3$ are only suspected and not confirmed to be produced by *E. coli*. Grey is a collection of carbonyls and hydrocarbons, with too many common spectral features to distinguish. Data from NIST[33]. See Supplementary Table 1 for details.

That such a simple organism as *E. coli*—and a single species at that—has a diverse enough metabolic machinery capable of producing a range of gases with useful spectral features is very promising for biosignature gas detection on exoplanets. While most of the gases are produced in small quantities on Earth there are exoplanet environments where the gases if produced in larger quantities could build up.

We showed previously that any biosignature gases produced can, under reasonable stellar EUV environments, survive photochemically in $H_2$-dominated atmospheres[32]. To emphasize that an $H_2$-rich atmosphere (i.e., with H radicals) can support accumulation of *E. coli*- and yeast-produced potential biosignature gases as compared to an oxic or oxidized atmosphere (with OH or O radicals) we compare radical reaction rates (from NIST[33]). We find that for all but one or two volatiles produced by *E. coli* or yeast, with reaction rates available at or close to room temperature, the H radical reaction rates are lower than the O-bearing radicals (OH and O) radical reaction rates (Supplementary Tables 3a, 3b, 3c). Therefore, if a gas can accumulate in an oxic or oxidized atmosphere it should also, in principle, be able to accumulate in an $H_2$-rich atmosphere.

## Summary

The main point of our experiment is to demonstrate that life can survive and thrive in an $H_2$-dominated atmosphere. We chose a 100% $H_2$ gas environment as a control, to show that life is viable even in the most extreme $H_2$ scenario. Atmospheres dominated by $H_2$ will always have other gas components that are products of planetary geology or atmospheric photochemistry. Furthermore, rocky planets will have to be colder than Earth, have a more massive surface gravity than Earth, and/or a replenishment mechanism to maintain an $H_2$-dominated atmosphere.

We used microorganisms that do not normally inhabit $H_2$-rich conditions, and at first glance might not be well adapted to $H_2$-rich atmospheres. Although *E. coli* is known to live in anaerobic conditions such as guts of many animals and has previously been studied in a pure $N_2$ environment[34], and yeast is commonly used in fermentation industry for beer, we have shown that both microorganisms are capable of surviving and reproducing in a 100% $H_2$ environment. The fact that both a simple cellular architecture like *E. coli* (a prokaryote) and much more complex single cellular microorganism like yeast (a eukaryote) can thrive in a pure $H_2$ gas environment, and produce a variety of byproduct gases, opens the possibility for a much broader spectrum of habitats for life on diverse habitable worlds.

**Acknowledgements:**

We thank A. Babbin for use of his laboratory and S. Smirga for assistance. We thank M. Slabicki and C. de Boer for providing us with a sample of yeast Saccharomyces cerevisiae S288C. We also thank J. Petkowska-Hankel for help with Fig. 1 and Z. Zhan for Fig. 4. Seed funding for this work came from the Templeton Foundation Grant 'The Alien Earths Initiative', ID 43769. Funding for this work came from the MIT Professor Amar G. Bose Research Grant Program.

**Author Contribution:**
SS conceived the original idea and wrote the paper with the help of JJP and MP. M.P. designed and implemented the experimental set-up. SS and MP planned the experiments with the help of JJP. JH and MP performed the experiments with the help of JJP. All authors analyzed the data.

**Author Information:**
Correspondence to: Sara Seager, Massachusetts Institute of Technology, 54-1718, 77 Mass. Ave. Cambridge, MA  02139; e-mail: seager@mit.edu

**Data Availability Statement:**
We supplied the source data for Figs. 2 and 3, which can be found as supplementary files at as at https://dspace.mit.edu/handle/ 1721.1/123824. The other data that support the plots within this paper and other findings of this study are available from the authors on request.

**Competing Interests:**
The authors declare no competing interests.